\documentclass[12pt]{article}
\usepackage[a4paper, total={7in, 10in}]{geometry}
\usepackage[parfill]{parskip}
\usepackage{physics, tensor, float, subcaption}
\usepackage{graphicx}
\graphicspath{ {Plots/} }
\usepackage{jhep-mod}
\usepackage{bm}
\usepackage{soul}
\usepackage{amssymb,amsmath,amsthm}
\usepackage{mathrsfs}
\usepackage[utf8]{inputenc}
\usepackage{enumerate}
\usepackage{bigints}
\usepackage{xcolor}
\usepackage{appendix}
\usepackage{graphicx}
\usepackage{float}
\usepackage{tikz}
\usepackage{setspace}
\usepackage{cancel}
\usepackage{doi}
\definecolor{purple}{rgb}{1,0,1}
\definecolor{lime}{HTML}{A6CE39} 

\definecolor{lime}{HTML}{A6CE39}
\newcommand{\orcidicon}{%
	\begin{tikzpicture}
	\draw[lime, fill=lime] (0,0) 
		circle [radius=0.16] 
		node[white] {{\fontfamily{qag}\selectfont \tiny ID}};
	\draw[white, fill=white] (-0.0625,0.095) 
		circle [radius=0.007];
	\end{tikzpicture}
	\hspace{-5mm}
}
\newcommand\orcidStefano{{\href{https://orcid.org/0000-0002-7632-7443}{\orcidicon}}}
\newcommand\orcidMatt{{\href{https://orcid.org/0000-0003-1088-6485}{\orcidicon}}}

\newcommand{\C}{\mathcal{C}}

\begin{document}


\title{\vspace{-25pt}\huge{
Painlev\'e--Gullstrand coordinates \emph{versus} Kerr spacetime geometry
}}


\author{
\Large
Matt Visser$\,^{1}$\!\orcidMatt
{\sf  and} 
Stefano Liberati$\,^{2,3,4}$\orcidStefano}

\affiliation{
$^1$ 
School of Mathematics and Statistics, Victoria University of Wellington, 
\\
\null\qquad PO Box 600, Wellington 6140, New Zealand.}

\affiliation{
$^{2}$  
SISSA - International School for Advanced Studies, Via Bonomea 265, \\
\null\qquad 34136 Trieste, Italy
}
\affiliation{
$^{3}$ 
IFPU - Institute for Fundamental Physics of the Universe, Via Beirut 2, \\
\null\qquad 34014 Trieste, Italy
}
\affiliation{
$^4$ 
INFN Sezione di Trieste, Via Valerio 2, 34127 Trieste, Italy
}

\emailAdd{matt.visser@sms.vuw.ac.nz}
\emailAdd{liberati@sissa.it}

\abstract{
\vspace{1em}

We discuss the tension between the possible existence of Painlev\'e--Gullstrand coordinate systems \emph{versus} the explicit geometrical features of the Kerr spacetime; a subject of interest to Professor Thanu Padmanabhan in the weeks immediately preceding his unexpected death. We shall carefully distinguish \emph{strong} and \emph{weak} Painlev\'e--Gullstrand coordinate systems, and \emph{conformal} variants thereof, cataloguing what we know can and cannot be done --- sometimes we can make \emph{explicit global} statements, sometimes we must resort to  \emph{implicit local} statements. For the Kerr spacetime the best that seems to be achievable is to set the lapse function to unity and represent the spatial slices with a 3-metric in factorized \emph{unimodular} form; this arises from considering the Doran version of Kerr spacetime in Cartesian coordinates. We finish by exploring the (limited) extent to which this construction might possibly lead to implementing an ``analogue spacetime'' model suitable for laboratory simulations of the Kerr spacetime. 

\bigskip
\noindent
{\sc Date:} Tuesday 20 September 2022; \LaTeX-ed \today.


\bigskip
\noindent{\sc PhySH:} Gravitation

\bigskip
\noindent{\sc Dedication}: To the memory of Professor Thanu Padmanabhan

\bigskip
To appear in the Professor Thanu Padmanabhan memorial volume 
published by\\
General Relativity and Gravitation
}

\maketitle
\def\tr{{\mathrm{tr}}}
\def\diag{{\mathrm{diag}}}
\def\cof{{\mathrm{cof}}}
\def\pdet{{\mathrm{pdet}}}
\def\d{{\mathrm{d}}}
\def\L{{\mathcal{L}}}
\parindent0pt
\parskip7pt
\renewcommand{\C}{\mathcal{C}}
\newcommand{\arctanh}{\tanh^{-1}}
\newcommand{\n}{\nabla}

\clearpage
\null
\vspace{-75pt} 
\section{Introduction }

In late August 2021, just a few weeks before his unexpected passing, Professor Thanu Padmanabhan expressed an interest in the interplay between Painlev\'e--Gullstrand coordinates~\cite{painleve1,painleve2,gullstrand}, or more generally spatially-flat coordinate systems, and the Kerr spacetime geometry~\cite{Kerr, Kerr-Texas, kerr-newman, Kerr-history, kerr-intro, kerr-book, kerr-book-2}. 
Extracted (with very minor edits) from an e-mail from TP to MV dated 25 August 2021:
\begin{quote}
Many spacetimes admit a 1+3 foliation such that the spatial sections are flat. (This is, of course what happens in PG coordinates). Given the metric in an arbitrary coordinate system is there a geometrical condition (necessary/sufficient) for such a foliation to exist?

Since we still have shift and lapse there are 4 dof; and arbitrary metric has 6 dof. So we have ``only'' lost 2 dof.... 
So I would expect  wide class of spacetimes to allow this.

Once you are in such a foliation all the curvature scalars of 3 space vanishes. Writing them in terms of 4-d curvature plus $K_{ab}$ etc one can write down some necessary conditions but which seem difficult to reduce to simple geometrical features. 

Any literature reference will also be useful. 
\end{quote}
MV responded on 27 August 2021, and that response, (now greatly expanded and elaborated), is the ultimate basis for the extensive discussion below. The fundamental reason these considerations are so interesting is the tension between the relative tractability of those spacetime metrics that can be expressed in Painlev\'e--Gullstrand coordinates~\cite{Poisson-Martel, Faraoni:2020}, and the overwhelming pedagogical~\cite{Weinberg, MTW, Adler-Bazin-Schiffer, Wald, D'Inverno, Hartle, Carroll, Hobson, Poisson, Padmanabhan}, theoretical~\cite{Teukolsky:2014, Adamo:2014, Hehl:2014}, and astrophysical~\cite{Bambi:2011, Johannsen:2015, Reynolds:2013, Bambi:2017, Reynolds:2019, LISA:2020} importance of the Kerr spacetime. 

\section{Preliminaries }

We shall start by distinguishing \emph{strong} and \emph{weak} forms of the Painlev\'e--Gullstrand coordinate systems~\cite{painleve1,painleve2,gullstrand}, and then --- by extension --- \emph{conformal} versions of the (strong/weak) Painlev\'e--Gullstrand coordinate systems. 

\subsection{Strong Painlev\'e--Gullstrand coordinates}

We shall say that a coordinate system is of \emph{strong} Painlev\'e--Gullstrand form if the spacetime line element
can be written as
\begin{equation}
\d s^2 = - \d t^2 + \Big|\d \vec x - \vec v(t,\vec x) \; \d t\Big|^2.
\end{equation}
That is, the metric can be cast in the form
\begin{equation}
g_{ab} = \left[ \begin{array}{c|c}
-1 +  \{\delta^{ij} v_i(t,\vec x) v_j(t,\vec x)\} & -v_j(t,\vec x) \\ 
\hline -v_i(t,\vec x) & \delta_{ij}
\end{array} \right]. 
\end{equation}
Equivalently, for the inverse metric
\begin{equation}
g^{ab} = \left[ \begin{array}{c|c}
-1 & -v^j(t,\vec x) \\ 
\hline -v^i(t,\vec x) & \delta^{ij} - v^i(t,\vec x) \; v^j(t,\vec x)
\end{array} \right]. 
\end{equation}
Here $v^i(t,\vec x)= \delta^{ij} \, v_j(t,\vec x)$.
In the language of the ADM formalism~\cite{ADM1, ADM2} the lapse function, typically denoted $N(t,\vec x)$, is unity and the spatial 3-slices are flat. For this class of metrics all nontrivial aspects of the spacetime geometry have been shoe-horned into the shift vector $N_i(t,\vec x) = g_{ti}(t,\vec x) = - v_i(t,\vec x)$. The relative minus sign appearing herein, $N_i = - v_i$, is merely formal --- a historical accident ultimately due to differing conventions between the ADM and analogue spacetime communities. Note however, that there is an additional choice of sign implicit in the choice between ``ingoing'' and ``outgoing'' Painlev\'e--Gullstrand coordinates; a distinction equivalent to reversing the sign of the time coordinate.\footnote{For the sake of concreteness, let us connect this discussion with the specific case of Schwarzschild spacetime. In this case it is easy to recognize that the metric element in the usual Schwarzschild coordinates $(t_{\text{s}},r_{\text{s}})$ can be recast in the strong Painlev\'e--Gullstrand form by either sign in the coordinate transformation  
\begin{equation}
\label{eq:Schw_Painleve_Gullstrand}
\text d t = \text d t_{\text{s}} \pm \frac{|v|}{\left( 1 - v^2\right)} \text d r;
\quad \text d r=\text d r_{\text{s}}.
\end{equation}
Here $|v|=\sqrt{2M/r}$, and the $+$ sign corresponds to a black hole spacetime and the $-$ sign to its time reversal, i.e.~a white hole.
}

\enlargethispage{20pt} 
\emph{Algebraically} this strong Painlev\'e--Gullstrand form can be achieved \emph{if and only if} there exist two distinct 4-orthogonal 4-vectors --- a covariant vector $T_a=(-1;0,0,0)$ and a contravariant vector $F^a=(0; v^i)$, satisfying $T_a F^a=0$, such that the metric factorizes in the specific form:
\begin{equation}
g_{ab} = \eta_{cd} \left( \delta_a{}^c - T_a F^c \right) 
\left( \delta_b{}^d - T_b F^d \right). 
\end{equation}
Here as usual $\eta_{ab} = \diag(-1,1,1,1)$.
Expanding, one has
\begin{equation}
g_{ab} = \eta_{ab} + (\eta_{cd} F^c F^d) T_a T_b 
- T_a (\eta_{bc} F^c) - (\eta_{ac}  F^c) T_b. 
\end{equation}
Thence, since $(\delta_a{}^c - T_a F^c) (\delta_c{}^b + T_c F^b) = \delta^a{}_b$, for the inverse metric we have:
\begin{equation}
g^{ab} = \eta^{cd} \left( \delta_c{}^a + T_c F^a \right) 
\left( \delta_d{}^b + T_d F^b \right). 
\end{equation}
Expanding
\begin{equation}
g^{ab} = \eta^{ab} + (\eta^{cd} T_c T_d) F^a F^b + (\eta^{ac} T_c) F^b
+ F^a (\eta^{bc} T_c). 
\end{equation}
Note that we must then have $g^{ab} T_b = (1;v^i)$, whereas $g_{ab} F^b  = (\{\delta_{jk} v^j v^k\}; v_i)$. Thence
$g^{ab} T_a T_b = -1 = \eta^{ab} T_a T_b$, while
$g_{ab} F^a F^b =  \{\delta_{jk} v^j v^k\} = \eta_{ab} F^a F^b$. Thus $T_a$ is a timelike unit co-vector with respect to both $g^{ab}$ and $\eta^{ab}$, while $F^a$ is a spacelike vector with respect to both $g_{ab}$ and $\eta_{ab}$. 

Note the extremely strong constraint on the co-vector $T_a$. One has $T_a= (-1,0,0,0)$ which implies $T_{[a,b]}=0$; so in the language of differential forms one has $T = - \d t$ and $\d T =0$. It is this $\d T=0$ constraint that will ultimately prove problematic for Kerr spacetime. 

Known examples of such strong Painlev\'e--Gullstrand  behaviour are:
\begin{itemize}
\item All of Schwarzschild spacetime; $r>0$. (See for example~\cite{Poisson-Martel,Faraoni:2020} and~\cite{heuristic}. For Schwarzschild spacetime one has $\vec v = \mp \sqrt{2m/r}\;  \hat r$ for black holes/white holes respectively.
\item Most of Reissner--Nordstr\"om spacetime; the region $r\geq Q^2/(2m)$. 
For Reissner--Nordstr\"om spacetime one has $\vec v = \mp \sqrt{2m/r-Q^2/r^2}\;  \hat r$ for black holes/white holes respectively. This becomes imaginary for $r<Q^2/(2m)$.
Since (for the usual black hole situation $m>|Q|$) the region $r\leq Q^2/(2m)$ lies below the inner (Cauchy) horizon, this deep-core breakdown of the strong Painlev\'e--Gullstrand coordinates is not particularly worrisome. (See for example \cite{Faraoni:2020}.) 
Even for an extremal Reissner--Nordstr\"om black hole, $m=|Q|$, the horizon is at $r_H=m$ while the breakdown of Painlev\'e--Gullstrand  coordinates occurs at $m/2$, well below the (extremal) horizon.  Finally for the case of naked singularities, (when $m<|Q|$),  the 
breakdown of Painlev\'e--Gullstrand  coordinates at $r\leq Q^2/(2m)$ is itself ``naked'' and in principle visible from asymptotic infinity.

\item 
Spatially flat $k=0$ FLRW cosmologies, and Kottler (Schwarzschild-de Sitter) spacetimes~\cite{PG-cosmo}. 

\item
The Lense--Thirring (slow-rotation Kerr) spacetime~\cite{Lense-Thirring, Pfister} also has a strong Painlev\'e--Gullstrand implementation~\cite{PGLT1, PGLT2, PGLT3, PGLT4}.
\item 
This also works (essentially by definition) for the generic Natario class of warp-drive spacetimes 
    (including the Alcubierre, zero-expansion, and zero-vorticity warp drives)~\cite{Santiago:2021-a, Santiago:ADM-warp}. 
\item 
This also works (essentially by definition) for the tractor/pressor/stressor beam spacetimes~\cite{Santiago:2021-b,Santiago:2021-c}.

\end{itemize}

Known examples of spacetimes that are \emph{incompatible} with such strong Painlev\'e--Gullstrand  behaviour are:
\begin{itemize}
\item Kerr (and the Kerr--Newman) spacetimes. The non-existence proof is quite tricky and indirect, using asymptotic peeling properties of the 3-dimensional  Cotton--York tensor~\cite{Valiente-Kroon:2004a,Valiente-Kroon:2004b}.
The spatial 3-slices cannot even be made conformally flat, let alone Riemann flat. 
There is also evidence that the 3-metric characterizing the spatial slices cannot even be diagonalized without adverse effects on other aspects of the Kerr geometry~\cite{Baines:Darboux}. 
\item
The van den Broeck warp drive~\cite{VanDenBroeck:1999a,VanDenBroeck:1999b}, a warp drive variant wherein the spatial 3-slices are allowed to be conformally flat instead of being Riemann flat. 
\end{itemize}

\subsection{Weak Painlev\'e--Gullstrand coordinates}

We shall say that a coordinate system is of \emph{weak} Painlev\'e--Gullstrand form if the spacetime line element
can be written as
\begin{equation}
\d s^2 = - N(t,\vec x)^2 \; \d t^2 + \Big|\d \vec x - \vec v(t,\vec x) \; \d t\Big|^2.
\end{equation}
That is, the metric can be cast in the form
\begin{equation}
g_{ab} = \left[ \begin{array}{c|c}
-N(t,\vec x)^2 +  \{\delta^{ij} v_i(t,\vec x) v_j(t,\vec x)\} & -v_j(t,\vec x) \\ 
\hline -v_i(t,\vec x) & \delta_{ij}
\end{array} \right]. 
\end{equation}
Equivalently, for the inverse metric
\begin{equation}
g^{ab} = \left[ \begin{array}{c|c}
-1/N^2 & -v^j(t,\vec x)/N^2 \\ 
\hline -v^i(t,\vec x)/N^2 & 
\delta^{ij} - v^i(t,\vec x) \; v^j(t,\vec x)/N^2
\end{array} \right]. 
\end{equation}

In the language of the ADM formalism~\cite{ADM1, ADM2}  the lapse function $N(t,\vec x)$ is now allowed to be non-trivial, while the spatial 3-slices are still flat. 
\enlargethispage{20pt}

Finding a factorized form of the metric is now a little trickier --- by considering the sub-case $v_i(x)\to 0$ it becomes clear that it is useful to consider the matrix $(\eta_N)_{ab} = \diag\{-N^2;1,1,1\}$. 
Then \emph{algebraically} weak Painlev\'e--Gullstrand behaviour can be achieved \emph{if and only if} there exist two 4-vectors $T_a=(-N;0,0,0)$ and $F^a=(0; v^i/N)$ such that $F^a T_a =0$ and the metric factorizes in the following manner:
\begin{equation}
g_{ab} = (\eta_N)_{cd} \left(\delta_a{}^c - T_a F^c \right) 
\left(\delta_b{}^d - T_b F^d \right). 
\end{equation}
Expanding
\begin{equation}
g_{ab} = (\eta_N)_{ab} + \{(\eta_N)_{cd} F^c F^d\} T_a T_b
+ \{(\eta_N)_{ac} F^c\}  T_b + T_a \{(\eta_N)_{bc} F^c\}.
\end{equation}

For the inverse metric we now have
\begin{equation}
g^{ab} = (\eta_N)^{cd} \left(\delta_c{}^a + T_c F^a \right) 
\left(\delta_d{}^b + T_d F^b \right). 
\end{equation}
Expanding
\begin{equation}
g^{ab} = (\eta_N)^{ab} + \{(\eta_N)^{cd} T_c T_d\} F^a F^b 
+ \{(\eta_N)^{ac} T_c\} F^b
+ F^a \{(\eta_N)^{bc} T_c\}. 
\end{equation}
Here we define $(\eta_N)^{cd}= \diag\{-N^{-2};1,1,1\}$.

Thence $g^{ab} T_b = (1,v^i)/N$,  and similarly $g_{ab} F^b  = (\{\delta_{jk} v^j v^k\}; v_i)/N$. This implies
$g^{ab} T_a T_b = -1 = (\eta_N)^{ab} T_a T_b$, while
$g_{ab} F^a F^b =  \{\delta_{jk} v^j v^k\}/N^2 = (\eta_N)_{ab} F^a F^b$,
and $F^aT_a=0$. Thus $T_a$ is a timelike unit vector with respect to both $g^{ab}$ and $(\eta_N)^{ab}$, while $F^a$ is a spacelike vector with respect to both $g_{ab}$ and $(\eta_N)_{ab}$, and these two 4-vectors are 4-orthogonal. 

Note the extremely strong constraint on the co-vector $T_a$: We have $T_a= (-N,0,0,0)$ which implies $T_{[a,b]}=N_{[,a} T_{b]}/N$ whence $T_{[a,b} T_{c]}=0$; in the language of differential forms $T = - N \d t$ and $T\wedge \d T =0$. It is this $T\wedge \d T=0$ constraint that will ultimately prove problematic for Kerr spacetime.

Known examples of such weak Painlev\'e--Gullstrand behaviour include:
\begin{itemize}
\item
All strong  Painlev\'e--Gullstrand metrics are special cases of the weak form. 
\item
Analogue spacetimes in the eikonal (ray optics, ray acoustics) limit~\cite{LRR, Visser:1997, Barcelo:2000-BEC, Visser:unexpected, Barcelo:2001-normal-modes, Visser:2001-of/for, Jain:2007, Visser:2010, Liberati:2005}; where the lapse function is to be interpreted as the signal propagation speed, $N(t,\vec x)\to c_s(t,\vec x)$, and the shift vector is to be interpreted as \emph{minus} the\break 3-velocity of the medium, $N_i(t,\vec x) = g_{ti}(t,\vec x)= - v_i (t,\vec x)$. 

\enlargethispage{20pt}
\item Spherical symmetry: Most spherically symmetric spacetimes, even if time-dependent, can \emph{at least locally} be put in weak Painlev\'e--Gullstrand form~\cite{Nielsen:2005}. The major obstructions to putting spherically symmetric spacetimes into weak Painlev\'e--Gullstrand form are the possible existence of wormhole throats, (where the area coordinate $r$ would fail to be monotone), or situations where the Misner--Sharp quasi-local mass $m(r,t)$ is negative (where the shift vector would become imaginary)~\cite{Faraoni:2020}. 

In terms of the 
Misner--Sharp quasi-local mass one can write the spherically symmetric weak-Painlev\'e--Gullstrand metric in the form
\begin{equation}
g_{ab} = \left[ \begin{array}{c|c}
-N(r,t)^2\{1-2m(r,t)/r \}  & \mp N(r,t) \sqrt{2m(r,t)/r} \; \hat r_i 
\\[3pt] \hline 
\vphantom{\Big{|}}
\mp N(t) \sqrt{2m(r,t)/r} \; \hat r_i & \delta_{ij}
\end{array} \right].
\end{equation}
The inverse (contravariant) metric is
\begin{equation}
g^{ab} = \left[ \begin{array}{c|c}
-N(r,t)^{-2}  & \mp N(r,t)^{-1} \sqrt{2m(r,t)/r} \; \hat r_i 
\\[3pt] \hline 
\vphantom{\Big{|}}
\mp N(t)^{-1} \sqrt{2m(r,t)/r} \; \hat r_i & \delta^{ij} - \frac{2m(r,t)}{r} \; \hat r^i  \; \hat r^j 
\end{array} \right].
\end{equation}
This explicit form of the metric makes manifest the requirement for positive Misner--Sharp quasi-local mass.  

\enlargethispage{20pt}
In particular, any spherically symmetric spacetime violating the positive mass theorem cannot be put into weak-Painlev\'e--Gullstrand form. Conversely if the spacetime can be put into weak-Painlev\'e--Gullstrand form in the asymptotic region, then the spacetime must have positive ADM mass.

Finally note that even when one can prove the local \emph{existence} of  these weak-Painlev\'e--Gullstrand coordinates, actually \emph{finding} the relevant coordinate transformation might in practice be prohibitively difficult~\cite{PG-cosmo, Nielsen:2005}. 
\end{itemize}

Known examples that are \emph{incompatible} with such weak Painlev\'e--Gullstrand behaviour are:
\begin{itemize}
\item Kerr (and Kerr--Newman) spacetimes. (As for the strong Painlev\'e--Gullstrand form, the behaviour for the weak Painlev\'e--Gullstrand form is no better.)
\item
The van den Broeck warp drive~\cite{VanDenBroeck:1999a,VanDenBroeck:1999b}.
(As for the strong Painlev\'e--Gullstrand form, the behaviour
for the weak Painlev\'e--Gullstrand form is no better.)
\end{itemize}

\subsection{Conformal (strong/weak) Painlev\'e--Gullstrand coordinates}
\enlargethispage{30pt}
We shall say that a coordinate system is of \emph{conformal} Painlev\'e--Gullstrand form if the spacetime line element is conformal to (either strong or weak versions of) the Painlev\'e--Gullstrand line element.  Either 
\begin{equation}
\d s^2 = \Omega^2(t,\vec x) \left\{
- \d t^2 + \Big|\d \vec x - \vec v(t,\vec x) \; \d t\Big|^2
\right\},
\end{equation}
or
\begin{equation}
\d s^2 =  \Omega^2(t,\vec x) \left\{
- N(t,\vec x)^2 \; \d t^2 + \Big|\d \vec x - \vec v(t,\vec x) \; \d t\Big|^2
\right\}. 
\end{equation}
The existence of these \emph{conformal} Painlev\'e--Gullstrand line elements is a relatively weak constraint, though still enough to preclude the Kerr or Kerr--Newman spacetimes.
The factorization properties we saw for strong or weak Painlev\'e--Gullstrand situations continue to hold with only a minimal amount of conformal rescaling.

Known examples of this \emph{conformal} Painlev\'e--Gullstrand behaviour include:
\begin{itemize}
\item The McVittie spacetime can be put into \emph{conformal} weak Painlev\'e--Gullstrand form~\cite{PG-cosmo}. 
\item The van den Broeck warp drive~\cite{VanDenBroeck:1999a,VanDenBroeck:1999b}
corresponds to the special case of setting $\Omega(t,\vec x)N(t,\vec x)\to 1$
in the  \emph{conformal} weak Painlev\'e--Gullstrand line element.

That is
\begin{equation}
\d s^2 =  
- \d t^2 + 
\Omega^2(t,\vec x) \Big|\d \vec x - \vec v(t,\vec x) \; \d t\Big|^2. 
\end{equation}
\item
Analogue spacetimes in the wave propagation limit, (physical  optics, physical acoustics)~\cite{LRR, Visser:1997, Barcelo:2000-BEC, Visser:unexpected, Barcelo:2001-normal-modes, Visser:2001-of/for, Jain:2007, Visser:2010, Liberati:2005} are often of this form. Here the lapse function is to be interpreted as the wave propagation speed, $N(t,\vec x)\to c_s(t,\vec x)$, while the shift vector is to be interpreted as being proportional to \emph{minus} the (typically non-relativistic) 3-velocity of the medium, $N_i(t,\vec x) = g_{ti}(t,\vec x)= - \Omega(t,\vec x) \, v_i (t,\vec x)$. The presence of the overall conformal factor $\Omega(t,\vec x)$ has to do with the details of deriving the relevant wave equation (d'Alembertian), and is typically some function of the background density and pressure~\cite{LRR, Visser:2010}.
\end{itemize}

Known examples that are \emph{incompatible} with such behaviour are:
\begin{itemize}
\item Kerr (and Kerr--Newman) spacetimes. (As for the strong and weak Painlev\'e--Gullstrand forms, the behaviour for the conformal  Painlev\'e--Gullstrand form is no better.)
\end{itemize}

\subsection{Summary}

In view of the comments above, we shall now put some effort into seeing just how close we can get to putting the Kerr spacetime into (strong/weak/conformal) Painlev\'e--Gullstrand form. 
It is quite remarkable just how many physically interesting spactimes can be recast in (strong/weak/conformal) Painlev\'e--Gullstrand form,
and just how stubborn the astrophysically important Kerr spacetime is in simply refusing to be recast in this form.

\section{The river model of Kerr spacetime}

To develop our discussion we shall adapt the so-called ``river model'' of Kerr spacetime~\cite{Hamilton:2004}, which is based on the Doran coordinate system~\cite{Doran:1999}. 
See the discussion in references~\cite{kerr-intro,kerr-book}, and more recently~\cite{Baines:Ansatz}.
We shall soon see that (for the Kerr spacetime) the ``fluid'' in the ``river'' is not a \emph{perfect fluid}.

\subsection{General framework}
\enlargethispage{20pt}
Appealing to the ``river model''~\cite{Hamilton:2004},
 we aim  to establish the plausibility of generically writing the spacetime metric in the somewhat messier (non Painlev\'e--Gullstrand) form:
\begin{equation} 
\label{E:ansatz}
\d s^2 = \Omega(x)^2 \left\{ - \d t^2 + \left| \vphantom{\Big{|}} \d \vec x - \vec v(x) \left\{\d t - \vec u(x) \cdot \d \vec x\right\} \right|^2\right\}.
\end{equation} 
Here $x=(t,\vec x)$ and the two 3-vectors $\vec u(x)$ and $\vec u(x)$ are perpendicular.
We see that for $\Omega(x)\to 1$ and $\vec u(x)\to 0$ this is simply a standard (strong) Painlev\'e--Gullstrand metric, and (as we have seen above) this line element is sufficient to describe very many but not all physically interesting spacetimes.

Even for $\Omega(x)\neq 1$ and $\vec u(x)\to 0$ this is still of \emph{conformal} (strong) Painlev\'e--Gullstrand form. 
But, in view of the analysis by  Valiente-Kroon~\cite{Valiente-Kroon:2004a,Valiente-Kroon:2004b}, setting $\vec u(x)\to 0$ is \emph{not} sufficient for describing either the Kerr or Kerr--Newman spacetimes. 

On the other hand once we allow $\vec u(x)\neq 0$ (even with $\Omega(x)\equiv 1$) this \emph{is} general enough to describe the Kerr geometry. See (for example) references~\cite{Doran:1999, Hamilton:2004}. 
Once we additionally allow $\Omega(x)\neq 1$ then,
at least from the point of view of counting degrees of freedom, this this is general enough to describe describe an arbitrary spacetime.

To see how the counting argument works: Note that in four dimensions the metric $g_{ab}$ has ten algebraically independent components. Apply four coordinate conditions on your coordinate chart. This leaves six ``physical'' degrees of freedom. (Off-shell, before imposing the Einstein equations.) The ansatz has the required six degrees of freedom, three in the vector $\vec v(x)$, two more in the orthogonal vector $\vec u(x)$, and one remaining degree of freedom in in the conformal factor $\Omega(x)$.

Let us now rewrite the assumed line element (\ref{E:ansatz}) as:
\begin{equation} 
\d s^2 = 
g_{ab} \;\d x^a \;\d x^b 
=
\Omega^2 \;\eta_{ab} \;(\d x^a - \beta^a [\alpha_m\d x^m] )\;
(\d x^b - \beta^b [\alpha_n\d x^n] ).
\end{equation} 
That is
\begin{equation} 
g_{ab}
=
\Omega^2 \;\eta_{cd} \;(\delta^c{}_a - \beta^c \alpha_a )\;
(\delta^d{}_b - \beta^d \alpha_b ).
\end{equation} 

Here we have introduced two 4-orthogonal 4-vectors
\begin{equation} 
\alpha_m = (1; -u_i);
\qquad
\beta^a = (0; v^i);  
\qquad 
\alpha_m \beta^m = - u_i v^i =0. 
\end{equation} 
Expanding, we recover (\ref{E:ansatz})  
as asserted.
The utility of introducing the two 4-vectors $\alpha_m$ and $\beta^m $ is that it allows for a very simple specification of a suitable orthonormal co-tetrad and tetrad.
(That the existence of a globally defined co-tetrad, and its associated globally defined tetrad, is physically important is discussed in reference~\cite{Rajan:2016-global}.) 
Note that these 4-vectors $\alpha_m$ and $\beta^m $ are very close in spirit to  the 4-vectors $T_a$ and $F^a$ we introduced when discussing the strong and weak   Painlev\'e--Gullstrand forms of the metric. 
The key difference is that now
\begin{equation}
 T_a= (-1;0,0,0) \qquad \longrightarrow \qquad \alpha_m =   (1; -u_i),  
\end{equation}
with the overall sign flip just being a physically unimportant historical accident.
The non-trivial part of this new construction is the introduction of the 3-vector $u_i(t,\vec x)$.

Using these definitions let us now write the Kerr metric as
\begin{equation} 
g_{ab} = \eta_{AB} \; e^A{}_a \; e^B{}_b.
\end{equation} 
Then co-tetrad and tetrad can be taken to be 
\begin{equation} 
e^A{}_a  = \Omega\left\{\delta^A{}_a - \beta^A \alpha_a\right\};
\qquad
e_A{}^a  = \Omega^{-1}\left\{\delta_A{}^a + \alpha_A \beta^a\right\}.
\end{equation} 
Thence
\begin{equation} 
e^A{}_a \; e_A{}^b  = \left\{\delta^A{}_a - \beta^A \alpha_a\right\} \left\{\delta_A{}^b + \alpha_A \beta^b\right\}
= 
\delta_a{}^b.
\end{equation} 
It is the 4-orthogonality of $\alpha$ and $\beta$ that keeps the matrix inversion relating tetrad and co-tetrad simple.
The inverse metric is 
\begin{equation} 
g^{ab} = \eta^{AB} \; e_A{}^a \; e_B{}^b.
\end{equation} 
To be fully explicit \enlargethispage{30pt}
\begin{equation} 
g_{ab} = \Omega^2 \;\eta_{AB} \; \left\{\delta^A{}_a - \beta^A \alpha_a\right\} \; \left\{\delta^B{}_b - \beta^B \alpha_b\right\},
\end{equation} 
and
\begin{equation} 
g^{ab} = \Omega^{-2} \;\eta^{AB} \; \left\{\delta_A{}^a + \alpha_A \beta^a\right\}\;
\left\{\delta_B{}^b + \alpha_B \beta^b\right\}.
\end{equation} 

\subsection{Issues specific to Kerr spacetime}

Let us now deal with some issues of specific relevance to the Kerr spacetime, which we shall conveniently write here in its Doran form 
\begin{equation}
    \d s^2=-\d t^2+\left[\frac{\rho\d r}{R}+\frac{\beta R}{\rho}\left(\d t-a\sin^2\theta \d\phi\right)\right]^2+\rho^2\d\theta^2+R^2\sin^2\theta \d \phi^2\,.
    \label{eq:Kerr-Doran}
\end{equation}
Here the quantities $\beta$, $R$, $\rho$, and $r$, are ultimately functions of the coordinates $\{x,y,z\}$, and the parameters  $\{m,a\}$. Explicitly
\begin{equation} 
\beta = {\sqrt{2mr}\over R};
\qquad
R = \sqrt{r^2+a^2};
\end{equation} 
and
\begin{equation} 
\rho = \sqrt{ r^2 + a^2\cos^2\theta};
\qquad
\cos\theta = {z\over r};
\end{equation} 
while
\begin{equation} 
r^4 - r^2(x^2+y^2+z^2-a^2) - a^2 z^2 = 0.
\end{equation} 
Finally, note that the Doran time and azimuthal angle used in the above metric are related to those of the standard  Boyer--Lindquist coordinates $(t_{BL},\phi_{BL})$ via the simple relations
\begin{eqnarray}
t&=&t_{BL}-\int_{r}^{\infty}{\frac{\beta \d r}{1-\beta^2}}\\[3pt]
\phi&=&\phi_{BL}=\phi-a \int_{r}^{\infty}{\frac{\beta \d r}{R^2(1-\beta^2)}}\,.
\end{eqnarray}

\subsubsection{Flow and twist vectors}

In their ``river model of black holes'' paper,  
(their equation (30) and thereafter), Hamilton and Lisle~\cite{Hamilton:2004}  describe  how to recast the above line element Eq.~\eqref{eq:Kerr-Doran} into the form
\begin{equation} 
\d s^2 = 
g_{ab} \;\d x ^a \;\d x^b 
=
\eta_{ab} (\d x^a - \beta^a [\alpha_m\;\d x^m] )\;
(\d x^b - \beta^b [\alpha_n\;\d x^n] ),
\end{equation} 
with
\begin{equation} 
\alpha_m = (1; -u_i);
\qquad
\beta^a = (0; v^i);  
\qquad 
\alpha_m \beta^m = - u_i v^i =0. 
\end{equation} 
Thence
\begin{equation} 
\d s^2= 
g_{ab} \;\d x ^a\; \d x^b 
=
- \d t^2 + \sum_i \left\{\d x^i - v^i [\d t - u_j \d x^j] \right\}^2.
\end{equation} 
That is, for the Kerr spacetime they can get away with setting $\Omega\to 1$.
Explicitly, (in the Cartesian form of the Doran coordinates that they adopt), they report
\begin{equation} 
u_i =\left( {ay\over R^2}, -{ax\over R^2}, 0 \right)
=
{a\over R^2}(-y,x,0),
\end{equation} 
and
\begin{equation} 
v^i = {\beta R\over\rho} \left( -{xr \over R\rho} , - {yr\over R\rho},
-{zR\over r\rho} \right)
=
 {\beta r\over\rho^2} \left( -x , - y,
-{zR^2\over r^2} \right).
\end{equation} 
Here $v^i$ is what they call the ``\emph{flow}'' vector and $u_i$ is what they call the ``\emph{twist}'' vector.
( The relation of this twist vector to vorticity being not entirely clear --- more on this point below.)
Note $v^i u_i=0$, in concordance with our general discussion above.

Note that the twist vector $u_i$ is independent of the mass parameter $m$, it just depends on the spin parameter $a$. Moreover, the only place that the mass parameter $m$ shows up is in the quantity $\beta$, which is a pre-factor in the flow vector $u^i$. 
For the 4-vectors $\alpha_m$ and $\beta^m $, and so implicitly the ``twist'' and the ``flow'',  for the Kerr spacetime one has
\begin{equation}
\alpha_m = \left( 1, -{ay\over r^2+a^2}, {a x \over r^2+a^2}, 0\right);
\end{equation}
and
\begin{equation}
\beta^m = \sqrt{2m r\over r^2+a^2} \;\; {r\over r^2+a^2\cos^2\theta} \;\;
\left( 0, -x, 
-y, 
-z {(r^2+a^2)\over r^2}\right).
\end{equation}

To be fully explicit, it is sometimes useful to observe:
\begin{equation} 
r^2 = {\sqrt{(x^2+y^2+z^2-a^2)^2 +4a^2z^2} +(x^2+y^2+z^2-a^2)
\over 2} ;
\end{equation} 
\begin{equation} 
{1\over r^2} ={\sqrt{(x^2+y^2+z^2-a^2)^2 +4a^2z^2} -(x^2+y^2+z^2-a^2)
\over 2 a^2 z^2} ;
\end{equation} 
while
\begin{equation} 
R^2 ={\sqrt{(x^2+y^2+z^2-a^2)^2 +4a^2z^2} +(x^2+y^2+z^2+a^2)
\over 2} ;
\end{equation}
\begin{equation} 
\qquad
{1\over R^2} =-{\sqrt{(x^2+y^2+z^2-a^2)^2 +4a^2z^2} -(x^2+y^2+z^2+a^2)
\over 2 a^2(x^2+y^2)} ;
\end{equation}
and
\begin{equation} 
\rho^2 = r^2+ {a^2 z^2\over r^2}  
= 
\sqrt{(x^2+y^2+z^2-a^2)^2 +4a^2z^2}.
\end{equation} 

Then for the Kerr solution explicit computation yields $\nabla\cdot \vec u=0$, so the twist vector is at least solenoidal (divergence free).
Unfortunately the ``vorticity'' $\nabla\times \vec u\neq 0$ is both non-zero and rather messy. Similarly the helicity $\vec u \cdot (\nabla\times \vec u)\neq 0$ and Lamb vector $\vec u \times (\nabla\times \vec u)\neq 0$ are both non-zero and quite messy. 
In view of the axial symmetry and the solenoidal condition on the twist, there must be a vector potential for the twist of the form
\begin{equation} 
\vec u = \nabla \times \left(0,0,\Psi(x,y,z,a)\right)
= \nabla \Psi(x,y,z,a) \times (0,0,1),
\end{equation} 
with $\Psi(x,y,z,a)$ a somewhat messy axisymmetric function of its variables. 

Similarly one can calculate both  $\nabla\cdot \vec v\neq 0$ and $\nabla\times \vec v\neq 0$ for the flow vector $\vec v$, though now both of these quantities are nonzero.
Finally, the extension to Kerr--Newman spacetime is straightforward --- in the metric/line-element one simply replaces $m \to m - {1\over2} Q^2/r$ in the function $\beta$, no other changes are required. 

In summary, the ansatz 
\begin{equation} 
\d s^2 = - \d t^2 + \left| \vphantom{\Big{|}} \d \vec x - \vec v(x) \{\d t - \vec u(x) \cdot \d \vec x\} \right|^2.
\end{equation} 
is sufficiently general to include the Kerr and Kerr--Newman spacetimes.
The spatial slices are however emphatically not flat, so they are not compatible with any of the usual formulations of Painlev\'e--Gullstrand coordinates, a point we shall expand on more fully below.

\subsubsection{Geometry of 3-space}

Let us now focus on the geometry of the spatial slices in Kerr spacetime.
Temporarily returning to our general ansatz, strip out the conformal factor ($\Omega\to 1$) and consider the simplified ansatz:
\begin{equation} 
\d s^2 = - \d t^2 + \left| \vphantom{\Big{|}} \d \vec x - \vec v(x) [\d t - \vec u(x) \cdot \d \vec x] \right|^2;
\qquad
\vec v \cdot \vec u =0.
\end{equation} 
Expand:
\begin{equation} 
\d s^2 = - \d t^2 + |v|^2 [\d t - \vec u(x) \cdot \d \vec x]^2
+ 2  [\d \vec x \cdot \vec v(x)] [\d t - \vec u(x) \cdot \d \vec x] 
+|\d \vec x|^2.
\end{equation} 
Defining $v_i = \delta_{ij} v^j$, one can read off the 3-metric:
\begin{equation} 
g_{ij} 
= \delta_{ij} + |v|^2 u_i u_j -  v_i u_j - u_i v_j
= \delta^{mn} (\delta_{im} - u_i v_m)  (\delta_{jn} - u_j v_n),
\end{equation} 
Note that only when the twist vector vanishes $\vec u \to 0$ does one regain flat spatial slices, while still keeping a nontrivial spacetime due to a nonzero flow.\footnote{One could also regain flat spatial slices by letting the flow vector vanishe $\vec v \to 0$, but that is uninteresting as it simply reduces to flat Minkowski space.}  
Note further that, because of the 3-orthogonality of the twist vector $u$ and flow vector $v$, we have $\det(\delta_{ij} - u_i v_j) = 1$; which implies $\det(g_{ij}) = 1$. 
Therefore the  3-metric is a volume preserving deformation of flat 3-space. 
So 3-space is not quite Riemann flat, but it is (in some sense) close.

One can also read off the ADM shift vector. It is now not just determined by the flow, we have an additional contribution from the twist:
\begin{equation} 
N_i = g_{ti} = -(v_i - |v|^2 u_i) =  - g_{ij} v^j \neq - \delta_{ij} v^j= - v_i.
\end{equation} 

Finally let us read off the ADM lapse function
\begin{equation} 
g_{tt} = -1 + |v|^2 =  -N^2 + g_{ij} v^i v^j = - N^2 + |v|^2.
\end{equation} 
Thence $N(x)=1$, this metric ansatz is (as previously advertised) unit lapse~\cite{Baines:unit-lapse}. 
Imposing unit lapse is useful --- it implies $g^{tt}=-1$, and then the covector field $V_a = -\partial_a t = (-1;0,0,0) $ is geodesic. 
Indeed $V^a = g^{ab} V_b$ is a future pointing 4-velocity geodesic 4-velocity  (unit 4-vector). 
There are quite a few unit-lapse representations of Kerr~\cite{Baines:unit-lapse}.
It turns out that for Kerr making $N\to1$ is relatively easy, 
while trying to enforce $g_{ij}\to\delta_{ij}$ is impossible.

Having now developed a solid grasp of the spatial and spacetime geometry, we shall turn to the possibility of mimicking Kerr spacetime by some sort of analogue model.


\section{Analogue models}

Historically, most (but certainly not all) ``analogue spacetimes'' were primarily built using (moving) perfect fluids --- and the perfect fluid condition implied that the spatial metric was isotropic, which forced the spacetime metric into (conformal) Painlev\'e--Gullstrand form for non-relativistic fluids. While it is true that for relativistic perfect fluids one can have somewhat more general metrics of the Gordon (disformal) form~\cite{Visser:2010,Fagnocchi:2010sn}, nonetheless the Kerr geometry has also proven impossible to accommodate within such a framework, except within specific, (sometimes rather nonphysical) limits, see e.g.~\cite{Giacomelli:2017eze,Liberati:2018uev,Liberati:2018osj}. 

In view of this  observation, if one wishes to develop an ``analogue model'' for Kerr based on some moving fluid then that moving fluid \emph{cannot} be a perfect fluid. 
Indeed, while the ``river model'' for Schwarzschild spacetime works relatively cleanly in terms of an isotropic fluid, (albeit a somewhat unphysical one), the fluid in the river model for a Kerr geometry has to be anisotropic. 
We do have examples of anisotropic fluids, such as \emph{liquid crystals}, but none of the standard liquid crystals seem to quite be appropriate for current purposes.

Let us recapitulate the key points: 
\begin{itemize}
\item The spatial metric is 
\begin{equation} 
g_{ij} 
= \delta_{ij} + |v|^2 u_i u_j -  v_i u_j - u_i v_j
= \delta^{mn} (\delta_{im} - u_i v_m)  (\delta_{jn} - u_j v_n),
\end{equation} 
where in an analogue spacetime context one would want to interpret $\delta^{mn}$ as the (contravariant) spatial metric seen by laboratory equipment, while $g_{ij}$ is the spatial metric seen by the analogue model.
\item The shift vector is
\begin{equation} 
N_i = -(v_i - |v|^2 u_i).
\end{equation}     
\item The lapse function is unity, $N=1$.
\end{itemize}

\medskip 
\enlargethispage{30pt}
From the above, the 3-metric ``factorizes''
\begin{equation} 
g_3 = (I -u\otimes v) \; (I -u\otimes v)^T.
\end{equation} 
Because of 3-orthogonality
\begin{equation} 
|v|^2 = \delta_{ij} v^i v^j = g_{ij} v^i v^j.
\end{equation} 
Up to a 3-d rotation one can enforce
\begin{equation} 
u_i \sim(0,u,0); \qquad v_i \sim(v,0,0); \qquad 
g_{ij} \sim \left[\begin{array}{ccc} 
1 & -uv &0\\-uv & \;\;1 +u^2 v^2\;\; &0\\0&0&1
\end{array}\right];
\end{equation} 
so we again see
\begin{equation}  
\det(g_{ij}) = 1.
\end{equation} 
So the anisotropic fluid in question mimics a volume-preserving deformation of flat Cartesian 3-space. 

The eigenvalues of the analogue metric with respect to the laboratory metric are determined by $\det(g_{ij}-\lambda \delta_{ij})=0$ and one finds
\begin{equation}  
\lambda \in \left\{ 1 ,\; 1+{u^2v^2\over2} \pm u v\sqrt{1+{u^2v^2\over4}} \right\}.
\end{equation} 
The three corresponding eigenvectors are
\begin{equation}  
(0,0,1); \qquad\left({uv\over2}-\sqrt{1+{u^2v^2\over 4}} ,\; {uv\over2} +\sqrt{1+{u^2v^2\over 4}},\; 0\right); \qquad (1,1,0).
\end{equation} 

So while the desired anisotropic fluid has relatively simple principal directions and eigenvalues, they depend on both flow and twist in a nontrivual manner --- there is no obvious physically constructable anisotropic fluid that would sucessfully mimic this 3-geometry --- it seems one is dealing more with a theoretical construction than an experimentally realizable one.

\section{Discussion} 

\enlargethispage{20pt}
Overall we can summarize the situation as follows: The strong Panilev\'e--Gullstand coordinate systems have two key features --- Riemann flat spatial slices ($g_{ij}\to \delta_{ij}$) and unit lapse ($N\to 1$). Historically most attempts at generalizing the Panilev\'e--Gullstand coordinate systems have focussed on keeping the spatial slices extremely simple (either flat or at worst conformally flat) while relaxing conditions in the lapse function $N(x)$. Unfortunately while the weak Panilev\'e--Gullstand and conformal Panilev\'e--Gullstand coordinate systems cover a lot of territory, they are incapable of describing the astrophysically important Kerr spacetime or the physically important Kerr--Newman spacetime. Instead, to get to the Kerr or Kerr--Newman spacetimes one must take a different route --- keep the lapse function unity, $N=1$, but allow a controlled deformation of the 3-geometry. 
A factorized volume-preserving deformation of 3-space, wherein $g_3 = (I -u\otimes v) \; (I -u\otimes v)^T$ with $\vec u\cdot \vec v=0$, proves to be sufficient for describing the Kerr or Kerr--Newman spacetimes. More generally we have argued that any spacetime can with some effort be cast in the form 
\begin{equation} 
\label{E:ansatz2}
\d s^2 = \Omega(x)^2 \left\{ - \d t^2 + \left| \vphantom{\Big{|}} \d \vec x - \vec v(x) \left\{\d t - \vec u(x) \cdot \d \vec x\right\} \right|^2\right\};
\qquad \vec u\cdot \vec v =0.
\end{equation} 
This is as close as we have been able to get to formulating a general and complete answer to Professor Thanu Padmanabhan's enquiry of 25 August 2021.
We shall sorely miss his reply.

\clearpage
\section*{Acknowledgements}
MV was directly supported by the Marsden Fund, 
via a grant administered by the Royal Society of New Zealand.\\
SL acknowledges funding from the Italian 
Ministry of Education and  Scientific Research (MIUR)  
under the grant  PRIN MIUR 2017-MB8AEZ.

\bigskip
\hrule\hrule\hrule
\vspace{-5pt}


\end{document}